# Surface-induced linear magnetoresistance in antiferromagnetic topological insulator MnBi$_2$Te$_4$

X. Lei[1,2], L. Zhou[1], Z. Y. Hao[1], X. Z. Ma[1], C. Ma[1], Y. Q. Wang[1], P. B. Chen[1], B. C. Ye[1,2], L. Wang[1,3], F. Ye[1], J. N. Wang[2,*], J. W. Mei[1,3,†], and H. T. He[1,4,‡]

[1]*Department of Physics, Southern University of Science and Technology, Shenzhen, Guangdong, China*

[2]*Department of Physics, The Hong Kong University of Science and Technology, Clear Water Bay, Hong Kong, China*

[3]*Shenzhen Institute for Quantum Science and Engineering, Southern University of Science and Technology of China, Shenzhen, Guangdong, China*

[4]*Shenzhen Key Laboratory for Advanced Quantum Functional Materials and Devices, Southern University of Science and Technology, Shenzhen, Guangdong, China*

**Abstract.** Through a thorough magneto-transport study of antiferromagnetic topological insulator MnBi$_2$Te$_4$ (MBT) thick films, a positive linear magnetoresistance (LMR) with a two-dimensional (2D) character is found in high perpendicular magnetic fields and temperatures up to at least 260 K. The nonlinear Hall effect further reveals the existence of high-mobility surface states in addition to the bulk states in MBT. We ascribe the 2D LMR to the high-mobility surface states of MBT, thus unveiling a transport signature of surface states in thick MBT films. A suppression of LMR near the Neel temperature of MBT is also noticed, which might suggest the gap opening of surface states due to the paramagnetic-antiferromagnetic phase transition of MBT. Besides these, the failure of the disorder and quantum LMR model in explaining the observed LMR indicates new physics must be invoked to understand this phenomenon.

* phjwang@ust.hk
† meijw@sustech.edu.cn
‡ heht@sustech.edu.cn

Topological insulators (TIs) are a new class of quantum materials with insulating bulk, but gapless surface states protected by time-reversal symmetry (TRS) [1,2]. The topological surface states (TSSs) have a linear energy dispersion and exhibit helical spin texture near the Dirac point [3,4]. By introducing magnetism into TIs, TRS-breaking TSSs are expected to be gapped and quantum anomalous Hall effect (QAHE) arises, which has spin-polarized chiral edge states even in the absence of magnetic fields [5]. QAHE was firstly realized in Cr-doped $(Bi, Sb)_2Te_3$ magnetic TI thin films in 2013 [6]. But due to the inhomogeneity or disorder induced by the random magnetic doping, the effect can only be observed at temperatures as low as several tens of mK. Therefore, it is of great interest to search for intrinsic magnetic TIs to obtain high-temperature QAHE. In this context, as a newly discovered intrinsic antiferromagnetic (AFM) TI, $MnBi_2Te_4$ (MBT) quickly attracts lots of attention [7-9].

MBT has a van der Waals layered structure similar to $Bi_2Te_3$, but with an extra Mn-Te layer inserted into the middle of each quintuple layer of $Bi_2Te_3$. Therefore, it consists of Te-Bi-Te-Mn-Te-Bi-Te septuple layers (SLs) stacking in the $c$ axis. Within each SL, the Mn moments are ferromagnetically correlated, with a perpendicular magnetic anisotropy. But due to the inter-layer antiferromagnetic coupling, MBT becomes an A-type AFM below Neel temperature ($T_N$). The value of $T_N$ is usually about 25 K and can be tuned by pressure [10,11]. Theoretical studies have shown that paramagnetic (PM) MBT above $T_N$ is a strong TI with gapless TSSs. But when the temperature is reduced below $T_N$, it becomes an AFM TI with gapped surface states due to the $\Theta\tau_{1/2}$ symmetry breaking, where $\Theta$ is the time-reversal symmetry and $\tau_{1/2}$ is the half translation operator connecting nearest spin-up and -down Mn atomic layers [7-9]. In the few-layer limit, MBT can exhibit the QAHE or axion insulator phase in zero fields, depending on the numbers of SLs [7-9]. Transport studies of few-layer MBT have confirmed the existence of these two intriguing phases in odd- and even-layer MBT films, respectively [12,13]. More interestingly, high-Chern-number and high-temperature QAHE were also discovered in MBT [14].

Although much progress has been made in this exciting field of MBT, open questions remain. According to the theoretical studies [7-9], the surface states breaking the $\Theta\tau_{1/2}$ symmetry should be gapped below $T_N$. Indeed, an ARPES study of MBT crystals below $T_N$ reveals such a surface state gap of about 70 meV [15]. The surface state gap opening is also critical to the observation of the QAHE and axion insulator phase in few-layer MBT [12,13]. But different from these results, latest ARPES studies show that the surface states are gapless regardless of the temperature (or magnetic phase) of MBT [16-18]. Therefore, whether the surface states of MBT are gapped below $T_N$ is still under debate. Besides this, it is also noticed that only bulk transport properties were reported in previous studies of thick MBT flakes or films, but without any transport signature of the surface states [19-21].

In this work, we have performed systematic magneto-transport study of thick MBT flakes with low electron density. Different from the negative magnetoresistance (MR) reported in previous studies [19,20], a positive linear magnetoresistance (LMR) is observed in high perpendicular fields up to 14 T. Tilted field measurement further reveals the two-dimensional (2D) nature of this LMR. A two-band analysis of the measured nonlinear Hall effect relates this phenomenon to the high-mobility surface states of MBT. Besides these, this LMR not only appears below $T_N$, but also persists up to at least 260 K, with a noticeable suppression of it near $T_N$. A detailed discussion about its physical origin rules out the widely adopted classical or quantum LMR model [22,23]. Our work demonstrates the existence of high-mobility surface states in thick MBT films and its manifestation as a 2D LMR in the transport properties of MBT. The suppression of LMR near $T_N$ might also suggest the surface state gap opening as MBT goes through the magnetic transition from PM to AFM at $T_N$.

The MBT single crystals studied in this work were synthesized by the flux method. High purity Mn (purity: 99.98%), Bi (99.999%), and Te (purity 99.999%) powder were mixed with the molar ratio of 1:10:16 and sealed in a quartz tube. The temperature of the mixture was then increased to 900°C in 9 hours. After kept at 900°C for 12 hours, the mixture was slowly cooled down to 590°C over 120 hours, at which temperature the excess flux was removed by centrifugation. Finally, the quartz tube was placed into water and cooled to room temperature. As shown in the upper inset of Fig. 1 (a), the cross-sectional high angle annular dark-field scanning transmission electron microscope (HAADF-STEM) image shows clearly the SL layered structure of MBT, indicating the single crystallinity of flux-grown MBT. Thick flakes (~100 nm) were mechanically exfoliated from these single crystals and then transferred onto Si substrates with a top layer of 280 nm $SiO_2$. Ar ion milling and e-beam lithography were implemented to pattern the flakes into six-terminal Hall bar structures for standard transport characterization, as shown in the inset of Fig. 4. Magneto-transport study of these Hall bar devices was performed in an Oxford TeslatronPT system with the magnetic field up to 14 T. A lock-in technique was applied to measure the magnetoresistivity. The amplitude and frequency of the measuring ac current is 5 µA and 17 Hz, respectively. Since similar transport behaviors were observed, we only show the results of a Hall bar device with the MBT thickness of 120 nm.

Fig. 1 (a) shows the temperature ($T$) dependence of resistance ($R$) of our MBT sample. The overall $R(T)$ curve shows a metallic behavior, different from that of few-layer MBT [12,13]. As seen in the lower inset of Fig. 1 (a), a resistance peak appears at about 20 K, indicating the antiferromagnetic transition of MBT, which has been reported in many previous studies of MBT [12-14, 19-21]. Thus, the Neel temperature ($T_N$) of our sample is 20 K, below which it becomes an A-type antiferromagnet.

Fig. 1 (b) shows the magnetoresistance (MR) curve of MBT measured at 1.6 K, with the magnetic field ($B$) applied perpendicularly to the flake surface. As the field increases, the MR curve shows a rapid increase at a characteristic field $B_{c1}$, due to the transition of MBT from the AFM phase to the canted AFM (CAFM) one. Further increasing the field leads to another transition into the ferromagnetic (FM) phase at $B_{c2}$, *i.e.*, the magnetic moments of Mn are polarized in the same direction as the field with $B > B_{c2}$. Note that these magnetic transitions have been discussed in previous transport studies of MBT [12-14,19-21]. Surprisingly, in the FM phase, the MR is positive and linearly dependent on the field from $B_{c2}$ to 14 T, as indicated by the dashed line in Fig. 1 (b). This is very different from the negative MR which was usually observed in MBT and ascribed to the spin disorder scattering [19,20]. We are also aware that a recent work reported similar positive unsaturated MR in MBT flakes [21]. But a systematic study of this phenomenon and an in-depth analysis of its physical origin is still lacking.

To gain more insight into this unsaturated linear MR (LMR), we have also studied it in tilted fields and at different temperatures. Fig. 2 (a) shows the MR measured at $T$=5 K, but with different tilting angles ($\theta$) of the field. As indicated in the inset of Fig. 2 (b), $\theta$ is defined as the angle between the surface normal and the field. As $\theta$ increases, the MR decreases. Especially, $B_{C2}$ shifts toward higher fields and the positive LMR observed above $B_{C2}$ weakens. At $\theta = 90°$, *i.e.*, the field is parallel with the current, only a weak negative MR is observed in high fields. To quantitatively characterize this tilting angle dependence of LMR, we linearly fit the normalized MR above $B_{C2}$ and plot the obtained slope of LMR ($k$) as a function of $\cos\theta$ in Fig. 2 (b). The linear dependence strongly indicates that the LMR only depends on the normal component of the field, thus revealing the two-dimensional (2D) nature of LMR.

Interestingly, this 2D LMR also appears with $T > T_N$. Fig. 3 (a) shows the MR curves at different temperatures. As the temperature is increased above $T_N$, the MBT flake becomes paramagnetic. Therefore, there is no more field-induced magnetic transitions from AFM to CAFM at $B_{C1}$ or from CAFM to FM at $B_{C2}$. Accordingly, the MR curve only displays a simple quadratic field dependence ($B^2$) in low fields (see the dashed fitting curve for $T=260$ K in Fig. 3 (a)). But in high fields, the LMR can still be observed and persists up to at least 260 K. It also exhibits the 2D nature, as evidenced by the linear dependence of $k$ on $\cos\theta$ at $T = 30$ K (see Fig. 2 (b)). From the MR data in Fig. 3 (a), we extract the slope $k$ and crossover field ($B_0$) of the LMR at different temperatures. The slope is determined by linearly fitting each normalized MR curve with $B > 10$ T and the crossover field is the field at which the MR curve deviates from the linear fitting curve by 0.5%. The temperature dependence of $k$ and $B_0$ is shown in Fig. 3 (b). Obviously, $k$ is enhanced with decreasing temperatures, although a suppression of it is noticed near $T_N$. We will discuss the physical implication of such a suppression later. As for the crossover field $B_0$, it shows a relatively weak temperature dependence with $T > T_N$. But when $T < T_N$, a rapid increase of it is observed, *i.e.*, the field range of LMR shrinks at low temperatures. It should be noted that the crossover field $B_0$ obtained with $T < T_N$ coincides with the critical field $B_{c2}$, at which the magnetic transition from CAFM to FM occurs (see Fig. 1 (b)). It is this magnetic transition that sets the lower bound for the field range of LMR when $T < T_N$.

There are mainly two physical models so far to account for the LMR observed in various systems [22,23]. The classical one based on inhomogeneity or disorder predicts $k \propto \langle\mu\rangle$ & $B_0 \propto \langle\mu\rangle^{-1}$ for $\langle\mu\rangle > \Delta\mu$, or $k \propto \Delta\mu$ & $B_0 \propto \Delta\mu^{-1}$ for $\langle\mu\rangle < \Delta\mu$, where $\langle\mu\rangle$ is the average mobility and $\Delta\mu$ is the mobility fluctuation due to disorder [22]. Besides this, Abrikosov proposed a quantum model for the LMR [23]. It predicts that in the quantum limit of a gapless or small gap system, *i.e.*, all the carriers condense into the lowest Landau level in high fields, an unsaturated LMR will emerge and the magnetoresistivity $\rho_{xx}$ can be described by the equation $\rho_{xx} \propto \frac{B}{n^2}$, where $n$ is the density of carriers [23]. Whether it is of classical or quantum origins, the LMR is closely associated with fundamental transport parameters, such as mobility and carrier density. This reminds us to further study the Hall effect of MBT.

We have performed the Hall measurement of MBT from 1.6 K to 260 K. For clarity, we only show in Fig. 4 (a) the Hall data obtained at 30 K and 260 K, respectively. The Hall resistivity ($\rho_{yx}$) clearly has a nonlinear field dependence, as can be seen from the comparison between the $\rho_{yx}(B)$ curve at 260 K and the straight dashed line in Fig. 4 (a). Nonlinear Hall effect is generally observed in TIs, since the topological surface state coexists with the bulk one, giving rise to two conducting channels in TIs [24-26]. It is thus natural to also observe this nonlinear Hall effect in MBT, which has been shown in theoretical studies [7-9] or experimental ARPES measurement [15-18] to possess both surface and bulk states simultaneously. We have used a two-band model to fit the Hall data, with the Hall resistivity described by

$$\rho_{yx}(B) = -\frac{B}{e} \frac{(N_S \mu_S^2/t + n_B \mu_B^2) + \mu_S^2 \mu_B^2 B^2 (N_S/t + n_B)}{(N_S \mu_S/t + n_B \mu_B)^2 + \mu_S^2 \mu_B^2 B^2 (N_S/t + n_B)^2} \qquad (1),$$

where $e$ is the electronic charge, $n_B$ and $\mu_B$ are the carrier density and mobility of the bulk states, $N_S$ and $\mu_S$ are the sheet carrier density and mobility of the surface states, and $t$ is the thickness of MBT flake. There also exists a constraint condition for the fitting parameters given by $\sigma(B = 0) = e(N_S \mu_S/t + n_B \mu_B)$, where $\sigma(B = 0)$ is the conductivity of MBT in zero magnetic field [24-26].

Fig. 4 (b) shows the extracted fitting parameters for the bulk and surface states, respectively. Considering the possible anomalous Hall contribution to $\rho_{yx}$ in the CAFM phase with $T < T_N$ [20,21], the two-band fitting is only applied to the Hall data above $T_N$. As can be seen in Fig. 4 (b), the bulk concentration $n_B$ exhibits a non-monotonic dependence on $T$. Similar behavior was also observed in a previous transport study of metallic MBT flakes [20]. But different from that study, our samples have much lower $n_B$ (~ $10^{19}$ cm$^{-3}$) and higher $\mu_B$ (~200 cm$^2$V$^{-1}$s$^{-1}$), about one order of magnitude lower and higher than the $n_B$ (~$10^{20}$ cm$^{-3}$) and $\mu_B$ (~ 40 cm$^2$V$^{-1}$s$^{-1}$) obtained in Ref. [20], respectively. As for the surface channel, both $n_S$ (or $N_S/t$) and $\mu_S$ increases with decreasing $T$. The surface mobility $\mu_S$ is at the order of 1000 cm$^2$V$^{-1}$s$^{-1}$, much higher than the bulk one $\mu_B$, consistent with the symmetry protected topological surface states [24-26].

The nonlinear Hall effect in Fig. 4 reveals the existence of a high mobility surface channel in our MBT sample. As shown in Fig. 2, the observed LMR in high fields only depends on the normal component of field, exhibiting a characteristic 2D nature. It is thus tempting to ascribe this 2D LMR to the surface states of MBT, similar to many previous studies of LMR in TIs [27-29]. Note that similar nonlinear Hall effect and LMR were also observed in an extrinsic AFM TI, *i.e.*, Sm-substituted $Bi_2Te_3$, both of which were regarded as transport signatures of topological surface states in addition to the anisotropic Shubnikov-de Hass oscillations [30].

With the $n_S$ and $\mu_S$ extracted from the nonlinear Hall effect, we can now discuss the possible mechanism for the LMR in MBT. According to the disorder model, $k \propto \langle\mu\rangle$ & $B_0 \propto \langle\mu\rangle^{-1}$ for $\langle\mu\rangle > \Delta\mu$, or $k \propto \Delta\mu$ & $B_0 \propto \Delta\mu^{-1}$ for $\langle\mu\rangle < \Delta\mu$ [22]. In either case, one can expect $k \propto B_0^{-1}$. But this is apparently not consistent with the results we obtain in Fig. 3 (b). The HAADF-STEM image in Fig. 1 (a) also shows the single crystallinity of our MBT flakes. Therefore, we can safely rule out this disorder model for LMR in MBT.

The quantum model has been adopted widely to explain the LMR observed in TIs [27-29], graphite [31], or small band gap semiconductors [32]. Although this model was originally proposed in the quantum limit, experimental studies have revealed that this model is also applicable to systems with several Landau levels occupied [27,32]. Nonetheless, Landau quantization is a prerequisite for this quantum model, *i.e.*, $\mu B > 1$. Since the surface mobility $\mu_S$ shown in Fig. 4 (b) is at the order of 1000 $cm^2V^{-1}s^{-1}$, the minimum magnetic field required for Landau quantization is 10 T. Although the LMR of our MBT sample indeed appears in high fields up to 14 T, the crossover field $B_0$ shown in Fig. 3 (b) is obviously much smaller than 10 T, indicating that the LMR begins to emerge in MBT in low fields without any formation of Landau levels. The observation of LMR up to 260 K in Fig. 3 further suggests the minor role of Landau quantization in this phenomenon. Besides these, the quantum model also predicts $\rho_{xx} \propto \frac{B}{n^2}$, or the LMR slope $\frac{\rho_{xx}}{B} \propto n^{-2}$ [23]. Based on the results in Fig. 3 and 4, we plot the LMR slopes obtained at different temperatures as a function of $N_S$ in Fig. 5. The slope indeed decreases with increasing $N_S$. But when we fit the data with the function $N_S^{-c}$, the best fitting yields $c \approx 3.1$, as indicated by the red curve in Fig. 5. Such a large deviation of $c$ from 2, as well as the minor role of Landau quantization, leads us to believe that the quantum model is also unlikely responsible for the LMR observed in our MBT samples. The failure of both the classical and quantum model in accounting for the 2D LMR in MBT calls for more theoretical studies of this phenomenon in the future.

Up to now, both negative and positive MR have been observed in high fields in MBT. The negative MR arises since magnetic Mn moments get polarized in high fields, suppressing the spin disorder scattering of carriers [19,20]. It is the bulk states of MBT that give rise to this negative MR. Therefore, this negative MR is usually observed in MBT samples with high bulk carrier densities (~ $10^{20}$ cm$^{-3}$) [19,20]. But in our work, from the two-band analysis of the nonlinear Hall effect in Fig. 4, the bulk carrier density is only about $10^{19}$ cm$^{-3}$, indicating that the Fermi energy of our sample is much lower. One can thus expect the more important role of surface states in the transport properties of our MBT samples. This might explain why positive LMR which is of a 2D character and ascribed to the high-mobility surface states is observed in our work. It is worth pointing out that a recent study also reveals a positive LMR in MBT flakes with the bulk carrier density at the order of $10^{19}$ cm$^{-3}$ [21]. But different from our interpretation above, the positive LMR is believed to arise from the bulk states of MBT and ascribed to the disorder model [21].

In Fig. 3 (b), a suppression of $k$ is noticed near $T_N$. Since the LMR of our sample is associated with the surface states as discussed above, such a suppression likely reflects a certain change in the surface states of MBT near $T_N$. In a previous study of TI Bi$_2$Se$_3$ films, it is found that the gap opening of the surface states can reduce the LMR in Bi$_2$Se$_3$ [29]. We thus suspect that the suppression of $k$ near $T_N$ might also suggest a gap opening of surface states when MBT endures the PM-AFM phase transition. This might contradict with recent ARPES studies of MBT, where gapless surface states are found to persist down to temperatures below $T_N$ [16-18]. But this speculation is in agreement with the theoretical prediction of AFM TI [7-9], and consistent with the successful realization of QAHE and the axion insulator phase in few-layer MBT, since the surface state gap opening is crucial to the occurrence of these phenomena [12,13]. Therefore, our work might provide indirect transport evidence for the surface state gap opening in MBT, or at least a certain change in the surface states, due to the PM-AFM phase transition at $T_N$.

In conclusion, we have studied the transport properties of MBT thick flakes with low carrier densities. An unsaturated linear magnetoresistance is observed in high fields and persists up to 260 K. Tilted magnetic field measurement also reveals the 2D nature of this phenomenon. The observed nonlinear Hall effect further suggests that this 2D LMR is associated with the high-mobility surface states of MBT. The failure of both the classical and quantum LMR model indicates that new physics must be invoked to understand this phenomenon. The suppression of LMR near the Neel temperature might also provide indirect evidence for the surface state gap opening in AFM TI below $T_N$.

ACKNOWLEDGEMENTS

This work was supported by the National Key Research and Development Program of China (No. 2016YFA0301703), the Science, Technology, and Innovation Commission of Shenzhen Municipality (No. KQJSCX20170727090712763 and No. ZDSYS20190902092905285), the program for Guangdong Introducing Innovative and Entrepreneurial Teams (No. 2017ZT07C062), and was in part supported by the Research Grants Council of the Hong Kong Special Administrative Region, China, grants 16301418 and C6013-16E.

**FIGURE**

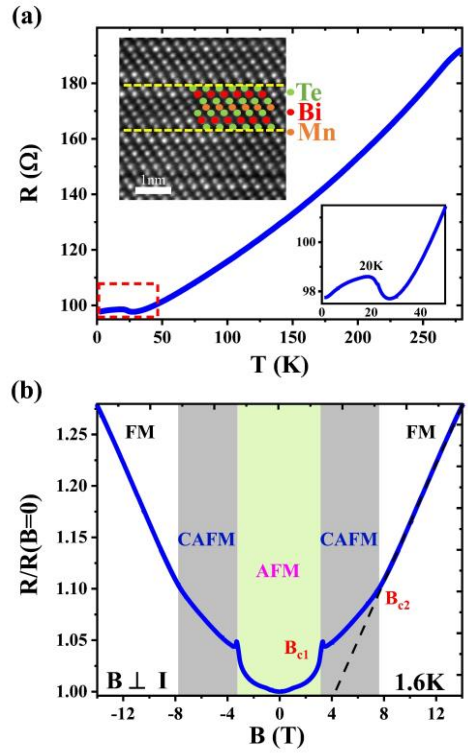

**Figure 1**

FIG 1. (a) Temperature dependent resistance of thick MBT flakes. Lower right inset: An enlarged view of the $R(T)$ curve enclosed by the dashed rectangle. Upper left inset: An HAADF-STEM image showing the SL layered structure of MBT with the scale bar of 1 nm.    (b) Normalized MR $R/R(B = 0)$ obtained in perpendicular fields and at $T$=1.6 K. Field-induced magnetic transitions from AFM to CAFM and then to FM occur consecutively at $B_{c1}$ and $B_{c2}$, respectively. The dashed line illustrates the linear field dependence of MR observed with $B > B_{C2}$.

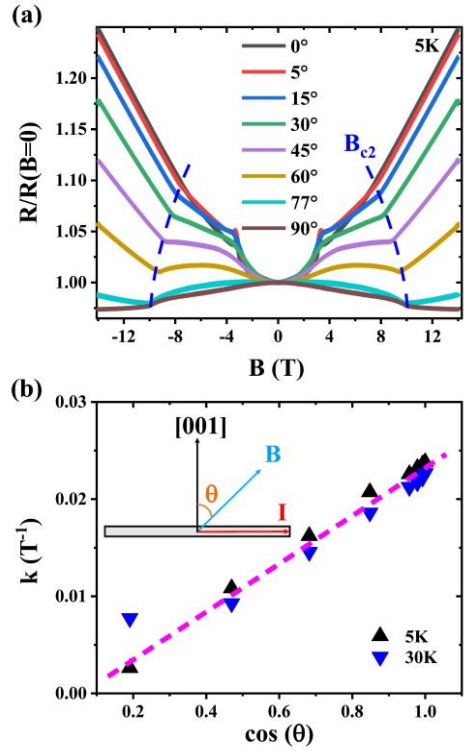

**Figure 2**

FIG 2. (a) Normalized MR measured at 5 K, but with different tilting angles $\theta$ as indicated. (b) The slope of normalized LMR $k$ as a function of $cos\theta$ at 5 K and 30 K, showing a linear dependence as illustrated by the dashed line. Upper left inset: Tilting angle $\theta$ defined as the angle between the surface normal and the field.

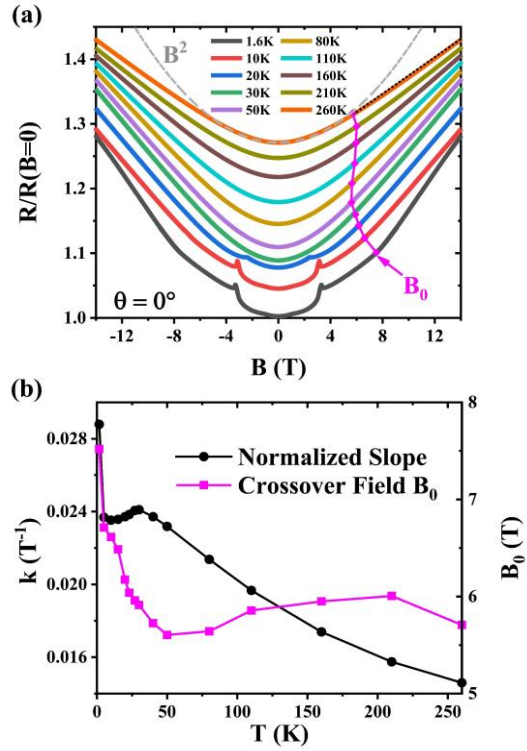

**Figure 3**

FIG 3. (a) Normalized MR obtained at different temperatures from 1.6 K to 260 K. At 260 K, the low-field MR show a quadratic field dependence, as indicated by the $B^2$ fitting curve. Also shown is the crossover field of LMR $B_0$ for each MR curve. (b) Temperature dependence of $k$ and $B_0$.

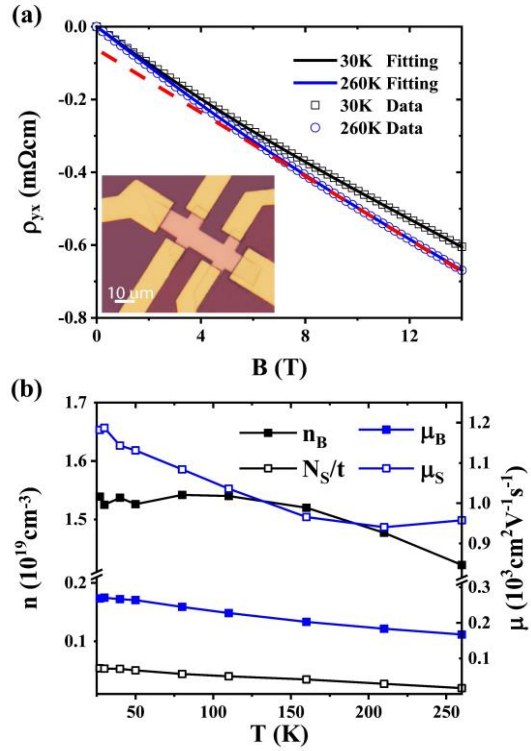

**Figure 4**

FIG 4. (a) Nonlinear Hall effect at 30 K and 260 K. The dashed line illustrates the nonlinearity of the obtained Hall resistivity. The Hall data can be well fitted by a two-band model, as indicated by the solid fitting curves. Inset: The photo of a six-terminal Hall bar device. (b) Fitting parameters for the bulk conduction band ($n_B$ & $\mu_B$) and surface states ($N_s$ & $\mu_s$) as a function of temperatures.

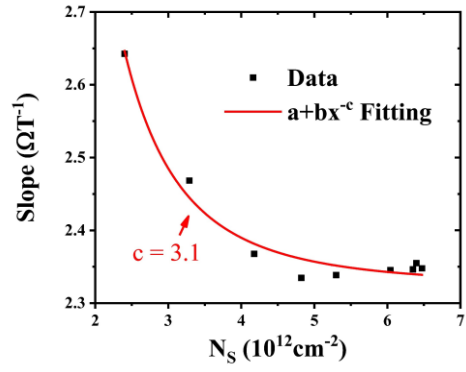

**Figure 5**

FIG 5. The surface sheet carrier density dependence of the LMR slope. Also shown is the best fitting curve with $c=3.1$.